\documentclass[12pt]{article}
\pdfoutput=1

\usepackage{indentfirst}
\usepackage{amsmath,amssymb,amsthm,amscd}
\usepackage[table]{xcolor}
\usepackage{graphicx}
\usepackage{slashed}
\usepackage{dsfont}
\usepackage{bbold}

\usepackage[backend=bibtex, sorting=none, style=numeric-comp]{biblatex}
\bibliography{mybib.bib}

\usepackage[body={6.5in, 9in}, right=1in, top=1in]{geometry}
\usepackage[format=hang,labelfont={bf},font={small,it}]{caption}
\usepackage[format=hang,labelfont={bf}]{subcaption}
\usepackage{booktabs}
\usepackage{multirow}
\usepackage{floatrow}
\DeclareFloatFont{tiny}{\tiny}
\definecolor{myBlue}{RGB}{15, 40, 220}
\usepackage[linktocpage=true]{hyperref}
\hypersetup{colorlinks=true, citecolor=myBlue, linkcolor=myBlue, urlcolor=violet}

 \newcommand{\C}{\mathcal{C}}

 \usepackage{soul}
 
\begin{document}
\def\thefootnote{\fnsymbol{footnote}}

\begin{center}
\Large{\textbf{Detecting Primordial $B$-Modes after Planck}} \\ \vspace{.5cm}

\large{Paolo Creminelli,$^{\rm a}$ Diana L\'opez Nacir,$^{\rm a,b}$ Marko Simonovi\'c,$^{\rm c,d,e}$\\ Gabriele Trevisan$^{\rm d,e}$ and Matias Zaldarriaga$^{\rm c}$}\\

\vspace{.5cm}

\small{
\textit{$^{\rm a}$ Abdus Salam International Centre for Theoretical Physics\\ Strada Costiera 11, 34151, Trieste, Italy}}
\vspace{.2cm}

\small{
\textit{$^{\rm b}$Departamento de F\'isica and IFIBA, FCEyN UBA, Facultad de Ciencias Exactas y Naturales, Ciudad Universitaria, Pabell\'on I, 1428 Buenos Aires, Argentina}}
\vspace{.2cm}

\small{
\textit{$^{\rm c}$Institute for Advanced Study, Einstein Drive, Princeton, NJ 08540, USA}}
\vspace{.2cm}

\small{
\textit{$^{\rm d}$ SISSA, via Bonomea 265, 34136, Trieste, Italy}}
\vspace{.2cm}

\small{
\textit{$^{\rm e}$ Istituto Nazionale di Fisica Nucleare, Sezione di Trieste, 34136, Trieste, Italy}}
\end{center}

\vspace{.3cm}
\hrule \vspace{.3cm}
\noindent \small{\textbf{Abstract}\\ 
We update the forecasts for the measurement of the tensor-to-scalar ratio $r$ for various ground-based experiments (AdvACT, CLASS, Keck/BICEP3, Simons Array, SPT-3G), balloons (EBEX 10k and Spider) and satellites (CMBPol, COrE and LiteBIRD), taking into account the recent Planck data on polarized dust and using a component separation method. The forecasts do not change significantly with respect to previous estimates when at least three frequencies are available, provided foregrounds can be accurately described by few parameters. We argue that a theoretically motivated goal for future experiments is $r\sim2\times10^{-3}$, and that this is achievable if the noise is reduced to $\sim1\,\mu$K-arcmin and lensing is reduced to $10\%$ in power. We study the constraints experiments will be able to put on the frequency and $\ell$-dependence of the tensor signal as a check of its primordial origin. Futuristic ground-based and balloon experiments can have good constraints on these parameters, even for $r\sim2\times10^{-3}$. For the same value of $r$, satellites will marginally be able to detect the presence of the recombination bump, the most distinctive feature of the primordial signal.
} 
\vspace{.3cm}
\noindent
\hrule
\def\thefootnote{\arabic{footnote}}
\setcounter{footnote}{0}
\vspace{.3cm}


\section{Introduction and motivations}
The year 2014 marked the beginning of the $B$-mode era in cosmology. After the direct detection of the lensing $B$-mode signal by Polarbear \cite{Ade:2014afa}, BICEP2 \cite{Ade:2014xna} pushed the constraints on primordial tensor modes using polarization to a level that is competitive with temperature. Given that temperature measurements are close to the cosmic-variance limit for the tensor-to-scalar ratio $r$, improvements in the future will practically only come from polarization. Planck \cite{Adam:2014oea} measured the level of polarized dust emission on the full sky with unprecedented precision (for previous measurements see for example \cite{Benoit:2003hg,Kogut:2007tq}). In this paper we want to look ahead at the future (and futuristic) experiments and understand whether the new data on dust polarization substantially change the reach expected for the various experiments.

 In looking at the future sensitivity on $r$, it is important to have in mind some motivated theoretical threshold. In the very near future we will explore the region $r \sim 0.1$, which corresponds to simple monomial potentials. If gravitational waves are not detected, is there another motivated threshold to reach? One might argue that $r \sim 2 \times 10^{-3}$ is a reasonable goal for future experiments. First of all, it approximately corresponds to the value predicted by potentials that approach asymptotically a constant as $\exp{(-\phi/M_{\rm P})}$ (Starobinsky model \cite{Starobinsky:1980te}, Higgs-inflation \cite{Bezrukov:2010jz}, etc.). A similar number is obtained by looking at the Lyth bound \cite{Lyth:1996im}. The excursion of the inflaton during inflation is given by
\begin{equation}
\frac{\Delta\phi}{M_{\rm P}} = \int {\rm d} N \sqrt{\frac{r}{8}} \;.
\end{equation}                                                                                                
If one assumes that $r$ monotonically increases going towards the end of inflation, one can conservatively replace $r(N)$ with the one on cosmological scales. The threshold $\Delta\phi = M_{\rm P}$ then corresponds to $r = 8 N^{-2} \simeq 2 \times 10^{-3}$. A detection of gravitational waves above this level would convincingly indicate a trans-Planckian displacement, under the mild assumption that $\epsilon$ increases as one moves towards the end of inflation. Another way to argue for the same threshold for $r$ is to study the consequences of imposing that the scalar tilt is of order $1/N$: $n_s-1 = -\alpha/N$ \cite{Creminelli:2014nqa,Mukhanov:2013tua,Roest:2013fha}. If this approximate equality is not an accident, but holds in a parametric window around $N=60$, one can argue for the existence of a forbidden region in $r$ between $10^{-1}$ and $10^{-3}$. This second number actually depends exponentially on the precise value of the scalar tilt, but $2 \times 10^{-3}$ corresponds to a reasonable lower bound within the present uncertainties on $n_s$  \cite{Creminelli:2014nqa}. All these theoretical prejudices should be taken with care, but motivate $2 \times 10^{-3}$ as a relevant figure of merit. Therefore, we will look ahead to check which experiments will be able to get to this value of $r$.\\

The outline is rather simple: in Section~\ref{Method} we explain the method used throughout this paper, while in Section~\ref{Results} we show the result obtained for various experiments. In Section~\ref{ResultsConservative} we consider more conservative analyses, focussing on possible evidences that the signal is indeed due to tensor modes.  
A similar study was done in Ref.~\cite{Lee:2014cya} concentrating on the superhorizon $B$-modes using \cite{Aumont:ESLAB}.

\section{Forecasting Method}\label{Method}
\subsection{CMB and Noise}
%
In linear perturbation theory, the coefficients $a_{\ell m}$ of the $T$-, $E$- and $B$-modes of the CMB are Gaussian random variables with zero mean and variance
\begin{equation}
\begin{split}
\langle a_{\ell m}^X a_{\ell' m'}^{Y*}\rangle=C_\ell^{XY} \delta_{\ell\ell'}\delta_{mm'},
\end{split}
\end{equation} 
where $X,Y=T,E,B$. From these, as customary, one defines the ``curly" correlators as
$\mathcal{C}_\ell^{XY}\equiv\ell(\ell+1)C_\ell^{XY}/(2\pi)$.
Due to parity invariance, only the $TT$, $EE$, $TE$ and $BB$ power spectra are necessary to characterize the CMB, the others being zero.  In our analysis we consider the $B$-mode power spectrum only, so we drop the superscript $BB$ where possible. This is generated by CAMB \cite{Lewis:1999bs} and, since we are solely interested in the forecast for the tensor-to-scalar ratio $r$, we set all cosmological parameters, except  $r$ and the optical depth $\tau$, to the current best fit values of Planck \cite{Ade:2015xua}.  Although this may look like a rough approximation, $r$ is expected to be only mildly degenerate with the other parameters, the biggest degeneracy  being the one with $\tau$ at low multipoles. We are going to marginalize over $\tau$ using a gaussian prior given by Planck analysis \cite{Ade:2015xua}. This is a conservative approach for satellites since they will have additional information on reionization. On the other hand, since large scale polarization measurements are affected by systematics, it is not clear how much they will improve the constraints on $\tau$. 

In the presence of white noise due to the instrumentation, the integration over a Gaussian beam to go from real space to harmonic space creates a bias $\mathcal{N}_{\ell}$ for the estimators of the power spectra. At each frequency channel, this is given by \cite{Knox:1995dq}
\begin{equation}
\begin{split}
\mathcal{N}_\ell=\frac{\ell(\ell+1)}{2\pi}\sigma_{\rm pix}^2 \,\Omega_{\rm pix}\, e^{\ell^2\sigma_b^2},
\end{split}
\end{equation}
where $\sigma_{\rm pix}$ is the noise variance per pixel of size $\Omega_{\rm pix}=\Theta_{\rm FWHM}^2$, and $\sigma_b=0.425\, \Theta_{\rm FWHM}$ is the beam-size variance. 
Our treatment here is clearly simplistic, since it does not take into account systematics. However, these are very experiment-dependent and go beyond the scope of this paper.
\subsection{Foregrounds}
The presence of foregrounds limits our ability in extracting the CMB signal from the data. Fortunately, each component scales differently in frequency, and thus it is possible to separate them using maps at different frequencies. In this paper we consider two diffuse components: synchrotron emission  ($S$) and thermal dust ($D$). 

The $BB$ power spectra of the Galactic synchrotron emission and thermal dust in antenna temperature is given by the following simple parametrization
\begin{equation}\label{eq:dust}
\begin{split}
S_{\ell,\nu}&=\left(W^S_\nu\right)^2 C_{\ell}^{S}=\left(W^S_\nu\right)^2 A_S \left( \frac{\ell}{\ell_S} \right)^{\alpha_S},\qquad W^S_\nu=\left( \frac{\nu}{\nu_S} \right)^{\beta_S}, \\
D_{\ell,\nu}&=\left(W^D_\nu\right)^2 C_{\ell}^{D}= \left(W^D_\nu\right)^2 A_D \left( \frac{\ell}{\ell_D} \right)^{\alpha_D},\qquad W^D_\nu= \left( \frac{\nu}{\nu_D} \right)^{1+\beta_D}
 \frac{e^{h \nu_D/kT} -1}{e^{h \nu/kT} -1},
\end{split}
\end{equation}
where the parameters are given in Tab.~\ref{tab:NewForegrounds}. This parametrization fits well the observed power spectra \cite{Page:2006hz,2014arXiv1409.2495P,Adam:2014oea}. Since for our analysis we are using as a reference the CMB signal, one has to rescale the antenna temperature of these components to the thermodynamic temperature of the CMB. The conversion is provided by the frequency dependence of the CMB
\begin{equation}
\begin{split}
W^{CMB}_\nu=\frac{x^2 e^x}{(e^x-1)^2}, \qquad x\equiv\frac{h \nu}{k\, T_{CMB}}.
\end{split}
\end{equation}
The frequency dependence of synchrotron and dust rescaled to the thermodynamic temperature of the CMB then reads
\begin{equation}\label{eq:TantTcmb}
\begin{split}
W^S_\nu\rightarrow W^S_\nu&=\frac{W^{CMB}_{\nu_S}}{W^{CMB}_\nu}\left( \frac{\nu}{\nu_S} \right)^{\beta_S}, \\
W^D_\nu\rightarrow W^D_\nu&=\frac{W^{CMB}_{\nu_D}}{W^{CMB}_\nu} \left( \frac{\nu}{\nu_D} \right)^{1+\beta_D}\frac{e^{h \nu_D/kT} -1}{e^{h \nu/kT} -1}.
\end{split}
\end{equation}

Synchrotron emission is the dominant one below 90 GHz\footnote{The synchrotron and dust power spectra are comparable at roughly 90 GHz for the cleanest 1\% of the sky. In regions with higher levels of polarized dust emission the transition happens at a lower frequency.},  while dust becomes increasingly important for higher frequencies. In our forecasts, for sky coverage of 70\% and 20\% we have fixed the synchrotron amplitude to the value measured by WMAP for the P06 mask \cite{Page:2006hz}, while for the 10\% and 1\% to the extrapolation of \cite{Flauger:2014qra} of the WMAP data. 
The synchrotron spectral index is known to steepen as Galactic latitude increases (see e.g.~\cite{Bernardo:2015aa}). However we checked that this effect can be safely neglected for our purposes (for a detailed study see \cite{Katayama:2011aa}).

\begin{table}[H]
\begin{center}
\begin{tabular}{c c c }
\toprule
{\bf Parameter} & {\bf Synchrotron} & {\bf Dust} \\
 \midrule
 $A_{72\%}$ [$\mu$K$^2$] & $2.1\times 10^{-5}$  & $0.169$\\
 $A_{53\%}$ [$\mu$K$^2$] & $2.1\times 10^{-5}$  & $0.065$\\
  $A_{24\%}$ [$\mu$K$^2$] & $2.1\times 10^{-5}$  & $0.019$\\
 $A_{11\%}$ [$\mu$K$^2$] & $4.2\times 10^{-6}$  & $0.013$\\
 $A_{1\%}$ [$\mu$K$^2$] & $4.2\times 10^{-6}$  & $0.006$\\
$\nu$ [GHz] &  65   & 353 \\
$\ell$ &  80 & 80\\
$\alpha$&$-2.6$ & $-2.42$\\
$\beta$ & $-2.9$ & 1.59\\
$T$ [K]&$-$ & $19.6$\\
\bottomrule
\end{tabular}
\caption{\em{Foreground parameters obtained or extrapolated from \cite{Page:2006hz,Adam:2014oea,Flauger:2014qra}, as explained in the text. $A_{f_{sky}}$ refers to the cleanest effective area $f_{sky}$.}}
\label{tab:NewForegrounds}
\end{center}
\end{table}
However, polarized dust emission is the leading contaminant for balloon and ground experiments probing frequencies higher than 90 GHz. A detailed measurement of the polarized dust has become available only recently \cite{Adam:2014oea}. For this reason, its impact has been underestimated in some previous analyses. For example, the observed value of the dust power spectrum at 353 GHz for $72\%$ of the sky is roughly 10 times bigger than the dust-model A used for the forecast of CMBPol \cite{Baumann:2008aq}. It is now clear that there are no regions in the sky for which a measurement of $r$ is achievable without having to deal with foregrounds \cite{Flauger:2014qra,Adam:2014oea}. One of the primary goals of this work is to  provide new forecasts for the detection and measurement of $r$ with more realistic levels of thermal dust contaminating the primordial signal. In this respect, we use the levels of dust presented in \cite{Adam:2014oea} for the $353$ GHz channel of Planck and extrapolate their results when needed. In particular, Planck \cite{Adam:2014oea} has recently provided the power spectra of dust at 353 GHz for a clean effective area of $72\%$, $53\%$, $24\%$, and $1\%$. For the 1\%-patch we use as a reference the value of the dust amplitude in the BICEP2 region. Even though Planck observed cleaner patches of the same size, the associated errors are too large to reliably determine the correct level of dust. 

Some experiments, e.g. Spider, will probe patches of the order of $10\%$ and we need to extrapolate the amount of dust on a patch roughly as big as theirs. This can be done in several ways. As a first guess, the interpolation of the values provided by Planck as a function of $f_{sky}$ gives $A_D^{BB}=0.013$. Another way consists in using the low $N_{\rm{HI}}$ region of \cite{2013arXiv1312.1300A} which covers $11\%$ of the sky. 
Using the relation $N_{\rm{HI}} = 1.41 \times 10^{26} \,\rm{cm}^{-2} \,\langle\tau_{353}\rangle$, where $\tau_{353}$ is the optical depth at 353 GHz, we find $N_{\rm{HI}}=1.35\times 10^{20}\,\rm{cm}^{-2}$. Substituting this value in the relation between the amplitude of $B$-modes and $N_{\rm{HI}}$ \cite{Adam:2014oea}, we find that $A_D^{BB}=0.013$. We tested this procedure against the amplitudes indicated by Planck for the $1\%$ of the sky for BICEP2, and found good agreement. A third way consists in using the relation between the amplitude  in polarization and the one in intensity  \cite{Adam:2014oea}, $A_D\propto I_{353}^{1.9}$. This relation for the same low $N_{HI}$ region gives $A_D^{BB}=0.009$. We will use $A_D^{BB}=0.013$ as an upper bound for the dust levels in the region observed by Spider.

An additional complication comes from the correlation between synchrotron and dust. It has been observed \cite{Flauger:2014qra} that the correlation among these two components can be as high as 70\%. To account for this effect in our forecast, for simplicity we will assume that in the power spectrum the correlation enters as $g \sqrt{S_{\ell,\nu_i} D_{\ell,\nu_j}}$ and set $g=0.5$ in our fiducial sky-model, independently of $f_{sky}$ and $\ell$.
\subsection{Likelihood and Fisher Analysis}\label{likeandfisher}

In the case of an experiment covering the whole sky, one can write the signal $d$ measured at the frequency $\nu_i$, in harmonic space, as
\begin{equation}
\begin{split}
d_{\ell,m}^{\nu_i}= \bar W^{\nu_i}_c a_{\ell,m}^c+n_{\ell,m}^{\nu_i}
\end{split}
\end{equation}
where $W$ provides the frequency dependence of each component\footnote{Since we are considering three components (the index $c$ runs over CMB, Dust and Synchrotron), and we are expressing everything in thermodynamic temperature, $W$ is a $3\times N_{\rm channel}$ matrix whose rows are $\left(1, W^D_{\nu_i},W^S_{\nu_i}\right)$.}, the bar indicates that the parameters are fixed to their ``true" value, and $n$ is the instrumental noise. 
Assuming that the amplitudes are Gaussian (also those of foregrounds), starting from the $\chi^2$  (and avoiding the use of indices)
\begin{equation}\label{Eq:chiCS}
\begin{split}
\chi^2 =\sum_{\ell,m}(d-W\cdot a)^{T}\cdot N^{-1}\cdot(d-W\cdot a)+  a^{T} \cdot C^{-1} \cdot a,
\end{split}
\end{equation}
where $C$ is the covariance matrix of the amplitudes of the various components and $N$ is the covariance matrix of the noise,  the likelihood can be written as
\begin{equation}
\mathcal{L}_{BB}(d,p) \propto\int {\rm d}^3 a \,e^{-\frac{1}{2}\chi^2}\propto
 e^{-\frac{1}{2}\sum_{\ell,m}d^T\cdot\left( W\cdot C\cdot W^T+N \right)^{-1}\cdot d}.
\end{equation}
In order to do a Fisher analysis we are interested in the average of the log-likelihood,
\begin{equation}\label{eq:likelihood}
\begin{split}
\langle \log{\cal L}_{BB}\rangle=-\frac{1}{2}\sum_{\ell} (2\ell+1)&\left[\log\text{Det}\left[
   \frac{ W\cdot{\C}_{\ell}^{BB}\cdot W^T+\mathcal{N}_\ell}{ \bar W\cdot\bar{\C}_{\ell}^{BB}\cdot \bar W^T+\mathcal{N}_\ell}\right]+\text{Tr}\left[
   \frac{\bar W\cdot\bar{\C}_{\ell}^{BB}\cdot\bar W^T+\mathcal{N}_\ell}{ W\cdot{\C}_{\ell}^{BB}\cdot W^T+\mathcal{N}_\ell}-{\rm I}\right]\right],
\end{split}
\end{equation}
where the normalization is such that for $\bar \C_\ell={\C}_\ell$, $\langle \log{\cal L}_{BB}\rangle=0$. In the following we will refer to the use of this formula as ``Component Separation" (CS).

Given the likelihood as a function of the parameters $\boldsymbol{p}$, one can define the Fisher matrix
\begin{equation}
F_{ij}= -\frac{\partial^2 \langle\log{\cal L}_{BB}\rangle}{\partial p_i\partial p_j}
 \Bigg|_{\boldsymbol{p}=\boldsymbol{\bar p}},
\end{equation}
where $\boldsymbol{\bar p}$ is the set of ``true" cosmological parameters.
The minimum error on the parameters given the data, is set by the Cramer-Rao bound to be
\begin{equation}\label{eq:sigma}
\begin{split}
\sigma_{p_i}^2\ge (F^{-1})_{ii}.
\end{split}
\end{equation}

In real experiments, only part of the sky is observed or can be used for cosmology, and thus not all modes are available for the analysis. To capture this effect in a simple way one can multiply the RHS of Eq.~(\ref{eq:likelihood}) by the fraction of the sky available $f_{sky}$.

In general the analysis is far more complicated. The likelihood as written in Eq.~(\ref{eq:likelihood}) becomes lossy, modes of multipole $\ell$ will be coupled with their neighbors. In addition, for experiments covering only a few percent of the sky, the leakage of $E$- into $B$-modes will add additional complications  to the detection of the signal produced by gravitational waves \cite{Smith:2006vq}. 
In our discussion, we will neglect these details, and thus the results produced with this method can be seen as optimistic. 

In the analysis of Sec. \ref{Results}, the likelihood in Eq.~(\ref{eq:likelihood}) is a function of 7 parameters\footnote{Notice that even in the limit in which there is no additional information coming from the difference in $\ell$-dependence, an experiment with three frequencies would allow a complete reconstruction of the parameters, since its covariance matrix would have six entries.} $\boldsymbol{p}=(r, A_D, A_S,\beta_D ,\beta_S, g, \tau)$, and the error on the tensor-to-scalar ratio is marginalized over the other parameters. Notice that  at low frequency the effect of changing $T_D$ is very similar to a change in $\beta_D$. As done in \cite{2014arXiv1405.0874P}, we are assuming that  the temperature of the dust emitting polarized radiation  is the same as the one determined from the intensity maps. In the following we will assume gaussian priors for $A_S$, $A_D$, $\beta_S$ with variance of 50\%, 50\%, 10\% respectively. For $\beta_D$ we use a gaussian prior of $10\%$, $30\%$ and $50\%$, for $f_{sky}\gtrsim50\%$, $\sim5-10\%$ and $\sim1\%$ respectively. For the optical depth we assume a gaussian prior given by Planck analysis \cite{Ade:2015xua}. No priors are assumed for the other parameters.

In Sec. \ref{ResultsConservative} we will instead take a more conservative approach and consider the likelihood in Eq.~(\ref{eq:likelihood}) as a function of 11 parameters. Namely, in addition to the 4 previous parameters, we will include the $\ell$-dependence of the foregrounds, $\alpha_D$ and $\alpha_S$, the spectral dependence of the CMB and its $\ell$-dependence. In particular we will constrain how much the CMB deviates from what expected using simple power laws with parameters $\alpha_{CMB}$ for the $\ell$-dependence and $\beta_{CMB}$ for the spectral dependence. Notice that $\alpha_{CMB}$ roughly corresponds to the tensor tilt.

In all the tables that follow, we use the symbol ``$-$" when $\sigma_r \geq r$.


\subsection{Likelihood and Fisher Analysis: a phenomenological approach}\label{sub:pheno}
The foreground model used in the previous section is of course approximate. For instance the foreground parameters may be space-dependent, and the distribution is not exactly gaussian. For this reason, and also to compare our results with previous forecasts, we also adopt a second, more phenomenological, method for estimating the error on the tensor-to-scalar ratio. It was proposed in \cite{Verde:2005ff} and already used also in previous forecasts for CMBPol \cite{Baumann:2008aq} and more recently also in \cite{Lee:2014cya}. The method assumes that with already known techniques such as \cite{Hobson:1998td,Baccigalupi:2002bh} it is possible to subtract the foregrounds up to a certain level in each single map. The noise introduced by the foreground removal is then modeled by accounting the number of possible cross correlations. The sum of foreground residual and noise is
\begin{equation}
\begin{split}
\mathcal{N}^{{\rm F}}_{\ell,\nu}=\sum_{F} \left( \sigma_{F}\,(\mathcal{S}_{\ell,\nu}+\mathcal{D}_{\ell,\nu})+\frac{4\,\mathcal{N}_{\ell, \nu_{tmp}}}{N_{\rm channel}(N_{\rm channel}-1)}\frac{W^F_\nu}{W^F_{\nu_{tmp}}}\right),
\end{split}
\end{equation}
where $\sigma_{F}$ is the fraction of leftover foregrounds  in power,  $W_\nu^F$ includes the conversion to thermodynamic temperature, $\nu_{tmp}$ is the reference frequency at which the foreground template has been created (e.g. $30$ GHz for synchrotron and $353$ GHz for dust), and $N_{\text{channel}}$ is the number of frequency channels. This quantity is then treated as an additional source of uncorrelated noise and as such is added to the noise bias. The resulting effective noise is given by
\begin{equation}\label{eq:effnoise}
\begin{split}
\left(\mathcal{N}^{eff}_\ell\right)^{-2}=\sum_{i,j\geq i}\left( \left(\mathcal{N}_{\ell, \nu_i}^F +\mathcal{N}_{\ell, \nu_i} \right) \left(\mathcal{N}_{\ell, \nu_j}^F +\mathcal{N}_{\ell, \nu_j} \right)\frac{1}{2}(1+\delta_{ij})\right)^{-1}.
\end{split}
\end{equation}
This method has the advantage of being independent on any specific technique of foreground subtraction, but, as already noted in \cite{Stivoli:2010rs} (see also \cite{2011PhRvD..84f3005E}), it has the downside of not being specific of any real experiment. By comparing the results of this and the CS method, one can estimate what is the level of foreground subtraction obtained by the component separation, and therefore the level at which the foreground modeling needs to be trusted. As already done in \cite{Verde:2005ff,Baumann:2008aq} and recently in \cite{Lee:2014cya}, in our forecasts we will assume that foregrounds can be subtracted at $1\%$ in power level in each map. 

The likelihood in this approach is the one of a single channel with an effective noise bias given by Eq.~(\ref{eq:effnoise}), and it reads
\begin{equation}\label{eq:likeBmodes}
\begin{split}
\langle\log{\cal L}_{BB}\rangle=-\frac{1}{2}\sum_{\ell} (2\ell+1)f_{sky}^{BB}&\left[\log\left(\frac{\C_{\ell}^{BB}+\mathcal{N}^{eff}_\ell}{\bar{\C}_{\ell}^{BB}+\mathcal{N}^{eff}_\ell}
 \right) + \frac{\bar{\C}_{\ell}^{BB}+\mathcal{N}^{eff}_\ell}{\C_{\ell}^{BB}+\mathcal{N}^{eff}_\ell} -1\right].
\end{split}
\end{equation}

In order to render the forecasts more realistic, we marginalize the error of the tensor-to-scalar ratio over the foreground residuals. This can be done by considering the percentage of foreground removal as an additional parameter and by multiplying the likelihood in Eq.~(\ref{eq:likeBmodes}) by a gaussian prior describing our ignorance about the exact level at which foregrounds have been removed. We will assume that the percentage of foreground removal $\sigma_F$ is known with relative error of order 1.

\subsection{Delensing}
One of the most important limiting factors in measuring primordial $B$-modes is gravitational lensing. Since $B$-modes induced by lensing have the same frequency dependence as primordial ones, one cannot proceed in the same way as with foregrounds. The idea then is to use the measurements of polarization or temperature on short angular scales to reconstruct the lensing potential and remove lensing effects from the $B$-mode polarization on the larger angular scales \cite{Knox:2002pe, 2002PhRvL..89a1304K, Seljak:2003pn, Smith:2010gu}. 

Polarization delensing is not always important in improving the constraints on $r$. In the case of a large tensor-to-scalar ratio, the dominant contribution to the signal-to-noise comes from the low multipoles. For example, the lensed $B$-modes power spectrum becomes comparable to the power spectrum of the primordial $B$-modes corresponding to  $r=0.1$ around $\ell\sim 100$. In this case delensing cannot improve the errors significantly. The other case in which delensing is irrelevant is when the noise is larger than the lensing signal. Indeed the power spectrum of lensing B-modes for $\ell\lesssim150$ is similar to white noise with amplitude $4.4\, \mu$K-arcmin, so that only for experiments with smaller noise delensing is relevant.
If the instrument has a high sensitivity and a good angular resolution needed for the reconstruction of the lensing potential, delensing the maps can become important and can improve the errors on $r$ even by a factor of a few for the beam sizes and the sensitivities of different proposed future experiments \cite{Seljak:2003pn,Smith:2010gu, 2012PhRvD..85h3006E}. For example,  Fig.~3 of \cite{Smith:2010gu} shows that with a beam size of 5 arcmin and sensitivity of $1\,\mu$K-arcmin one can improve the constraints on $r$ by a factor of 5. Moreover, as the sensitivity and the resolution increase, there are no limits in how much of the lensing signal can be subtracted. This improvement is marginal for all the experiments considered in this paper, except for the generation-IV experiments (GRD, BAL and ULDB) and for the satellites COrE and CMBPol. For all other experiments either the noise level is high enough to make delensing irrelevant or the angular resolution is too large to implement delensing. To include delensing in those experiments where it is viable, we assume that the power of lensing $B$-modes is reduced to 10$\%$ of the original value. The residual power would correspond to a noise equivalent power of $1.4\,\mu$K-arcmin. 
It would be useful to explore how the presence of foregrounds impacts the ability of delensing, but this goes beyond the scope of our paper.

\section{Results}\label{Results}
\subsection{BICEP2/Keck and Planck}

Let us begin our analysis by checking that our forecasting method is compatible with the recent recent joint analysis \cite{Ade:2015tva}. In order to do so, we combine the BICEP2/Keck 150 GHz channel with the Planck 353 GHz one. With the CS method, for a fiducial value of $r=0.05$ using the multipoles [20-150] and a gaussian prior on $\beta_D$ with variance $\sigma_{\beta_D}=0.11$, we obtain an error $\sigma_r=0.04$, which is in good agreement with what was reported in \cite{Ade:2015tva}.

\begin{figure}[htbp]
\begin{center}
\includegraphics[width=0.5 \textwidth]{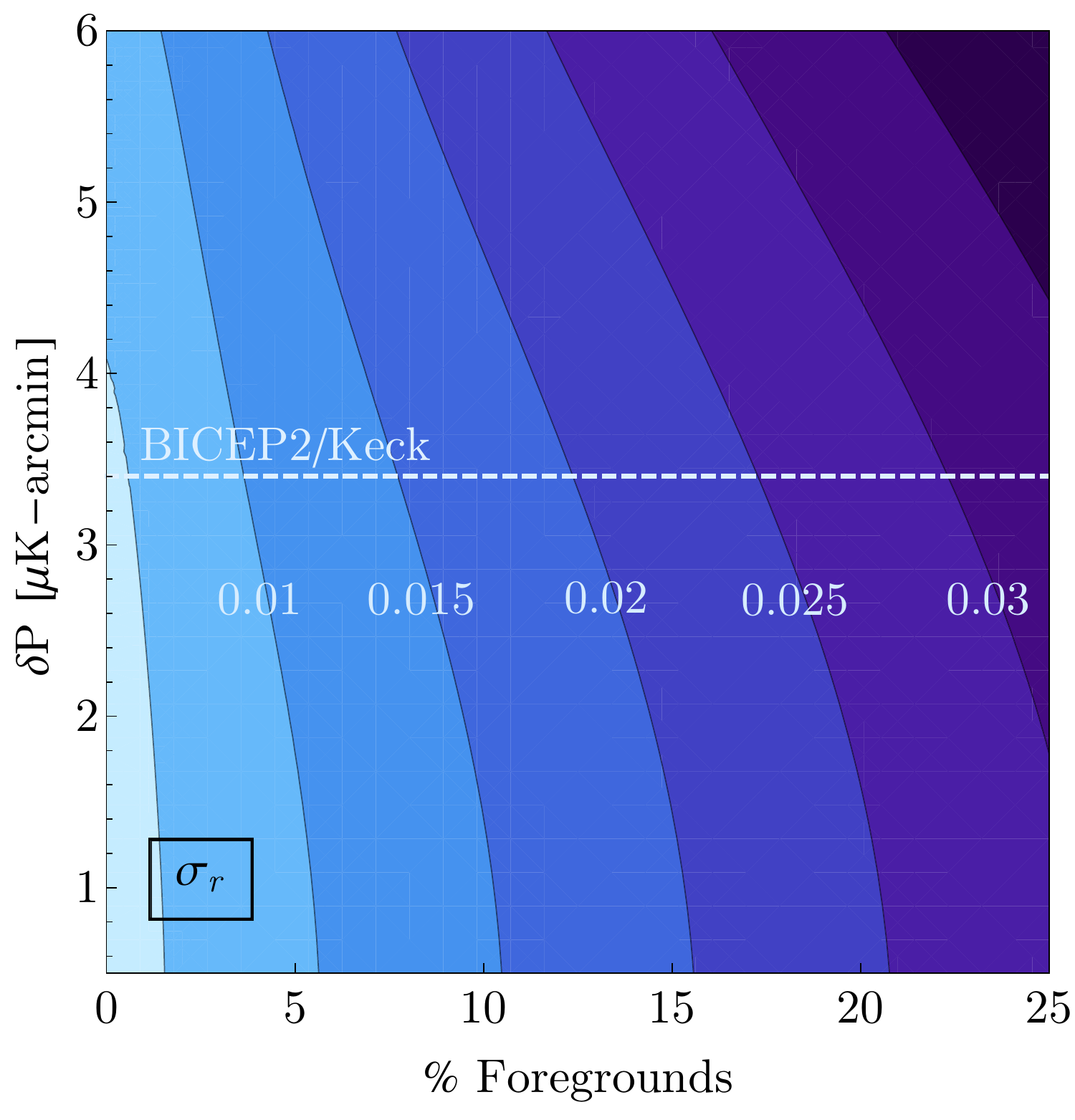}
\end{center}
\caption{ Contours represent the 1$\sigma$ error on $r$ for $r=0$, marginalized over the foreground residuals, using a single map of $1\%$ of the sky at $150 \,\rm{GHz}$ with beam of 30 {\rm arcmin}, as a function of foreground residuals and instrumental sensitivity. This is the case of BICEP2/Keck, which has sensitivity of 3.4 $\mu${\rm K-arcmin}.}
\label{fig:Biceplike}
\end{figure}

Alternatively, using the phenomenological approach of \ref{sub:pheno}, in Fig.~\ref{fig:Biceplike} we plot the contours representing the error on $r$ assuming $r=0$ as a function of sensitivity and foreground residuals.  As can be seen, our ability to constrain $r$ crucially depends on the foreground removal. A reduction of foregrounds to 10\% in power can lead (in the absence of gravitational waves) to a quite strong upper bound $r\leq0.05$ at $3\sigma$. Polarization experiments are already competitive with constraints from temperature alone, even with only one frequency and very mild foregrounds removal. Notice that with $T$-modes only cosmic variance prevents to constrain $r$ better than $5\times10^{-2}$ \cite{AML}.\footnote{This bound of course is impossible to achieve due to sample variance induced by masking the sky.}
Notice moreover that in the near future it will be possible to include in the analysis the 95 GHz channel of Keck (here we assume a noise of $9\,\mu$K-arcmin). As can be seen in Tab.~\ref{tab:PLANCKBICEP}, the inclusion of such a frequency would allow to reduce the error on the tensor-to-scalar ratio by a factor of 2 with respect to the current constraint.

\begin{table}[htb] 
\begin{center}
\begin{tabular}{ c c c}
\toprule
& {\bf r} & \bf BICEP2/Keck + $\boldsymbol{\rm{Planck}_{353}}$ \\
\midrule
\multicolumn{1}{ c }{\multirow{4}{*}{CS}} 		& 0.1 	& $3.5\times 10^{-2}$\\
\multicolumn{1}{ c }{} 													& 0.01 	& ---		\\	
\multicolumn{1}{ c }{} 													& 0.001 	& ---		\\			
\multicolumn{1}{ c }{} 													& 0 		& $2.2\times 10^{-2}$		\\			 	
\bottomrule
\end{tabular}
\caption{$1\sigma$ errors on $r$ for BICEP2/Keck  (95 and 150 GHz) and the 353 GHz channel of Planck. This error, as calculated from the phenomenological method of Sec.~\ref{sub:pheno}, corresponds to a 30\% level of leftover foregrounds in power.}
\label{tab:PLANCKBICEP}
\end{center}
\end{table}

\begin{table}[htb] 
\begin{center}
\begin{tabular}{ c c c c c c c}
\toprule
& {\bf r} & {\bf Keck/BICEP3} & {\bf Simons Array} & {\bf AdvACT}& {\bf CLASS} & {\bf SPT-3G}\\
\midrule
\multicolumn{1}{ c }{\multirow{4}{*}{CS}} 			& 0.1 	& $1.9\times 10^{-2}$ 	& $7.6\times 10^{-3}$	& $7.3\times 10^{-3}$	& $6.5\times 10^{-3}$	& $9.0\times 10^{-3}$\\
\multicolumn{1}{ c }{} 						& 0.01 	& $9.3\times 10^{-3}$ 	& $5.0\times 10^{-3}$	& $4.5\times 10^{-3}$	& $3.4\times 10^{-3}$	& $4.2\times 10^{-3}$\\
\multicolumn{1}{ c }{} 						& 0.001 	& ---				 	& ---					& ---					& ---	& ---\\
\multicolumn{1}{ c }{} 						& 0 		& $8.1\times 10^{-3}$ 	& $3.4\times 10^{-3}$	& $2.6\times 10^{-3}$	& $9.0\times 10^{-4}$		& $3.7\times 10^{-3}$\\
\midrule
\multicolumn{1}{ c }{\multirow{4}{*}{  FG 1\%}} 		& 0.1 	& $1.8\times 10^{-2}$ 	& $ 1.1\times 10^{-2}$	& $9.5\times 10^{-3}$	& $6.0\times 10^{-3}$		& $8.1\times 10^{-3}$\\
\multicolumn{1}{ c }{} 						& 0.01 	& $7.5\times 10^{-3}$ 	& $ 8.1\times 10^{-3}$	& $6.9\times 10^{-3}$	& $3.5\times 10^{-3}$		& $4.1\times 10^{-3}$\\
\multicolumn{1}{ c }{} 						& 0.001 	& --- 					& ---					& --- 					& ---						& ---\\
\multicolumn{1}{ c }{} 						& 0 		& $6.1\times 10^{-3}$ 	&$7.8\times 10^{-3}$		& $6.6\times 10^{-3}$	& $3.3\times 10^{-3}$		& $3.7\times 10^{-3}$\\
\bottomrule
\end{tabular}
\caption{ $1\sigma$ errors on $r$ for future ground-based experiments.}
\label{tab:ErrorGRD}
\end{center}
\end{table}

\subsection{Balloon-borne and Ground-based Experiments} \label{ResultsBal}
The situation will improve in the next few years since there are several experiments that are already collecting data and will have maps in two or more frequencies. In our forecasts we concentrate on a few proposed and funded experiments. In particular, for what regards ground-based experiments we consider Keck/BICEP3 and the Simons Array, and also AdvACT, CLASS and SPT-3G. The specifications used are in Tab.~\ref{tab:SpecBAL}. We vary the level of foregrounds  according to the fraction of the sky targeted by each experiment, as given in Tab.~\ref{tab:NewForegrounds}. For what concerns the available multipoles, we consider the range $[30,150]$. For AdvACT, CLASS and the Simons Array we consider the range $[2, 150]$, since they observe a larger fraction of the sky. This range is probably larger than what these experiments will  actually be able to observe, since to handle atmospheric contamination they need to filter the data, losing power at low $\ell$'s. In Sec.~\ref{ResultsConservative} we will take a more conservative perspective. As can be seen in Tab.~\ref{tab:ErrorGRD}, which summarizes our forecasts, we expect these experiments to explore values of $r$ of order $10^{-2}$.   The CS method gives results which are roughly comparable to a reduction of foregrounds to 1\%. Of course there are sometimes sizable differences, since we do not expect all experiments to reduce foregrounds in the same way. 
Still 1\% represents a rough estimate of the level at which one should trust the foreground modeling for the CS. 

\begin{table}[htb] 
\begin{center}
\begin{tabular}{ c c c c}
\toprule
& {\bf r} & {\bf EBEX 10k} & {\bf Spider} \\
\midrule
\multicolumn{1}{ c }{\multirow{4}{*}{CS}} 			& 0.1 	& $1.5\times 10^{-2}$ 	& $1.8\times 10^{-2}$\\
\multicolumn{1}{ c }{} 						& 0.01 	& $7.4\times 10^{-3}$				 	& ---	\\
\multicolumn{1}{ c }{} 						& 0.001 	& ---				 	& ---	\\
\multicolumn{1}{ c }{} 						& 0 		& $6.4\times 10^{-3}$ 	& $1.3\times 10^{-2}$\\
\midrule
\multicolumn{1}{ c }{\multirow{4}{*}{  FG 1\%}} 		& 0.1 	& $2.2\times 10^{-2}$ 	& $2.6\times 10^{-2}$\\
\multicolumn{1}{ c }{} 						& 0.01 	& --- 				& ---\\
\multicolumn{1}{ c }{} 						& 0.001 	& --- 				& ---\\
\multicolumn{1}{ c }{} 						& 0 		& $9.2\times 10^{-3}$ 	& $2.1\times 10^{-2}$\\
\bottomrule
\end{tabular}
\caption{ $1\sigma$ errors on $r$ for future balloon-borne experiments. }
\label{tab:ErrorBAL}
\end{center}
\end{table}
\begin{figure*}[t!]
    \centering
    \begin{subfigure}[t]{0.5\textwidth}
        \centering
        \includegraphics[width= \textwidth]{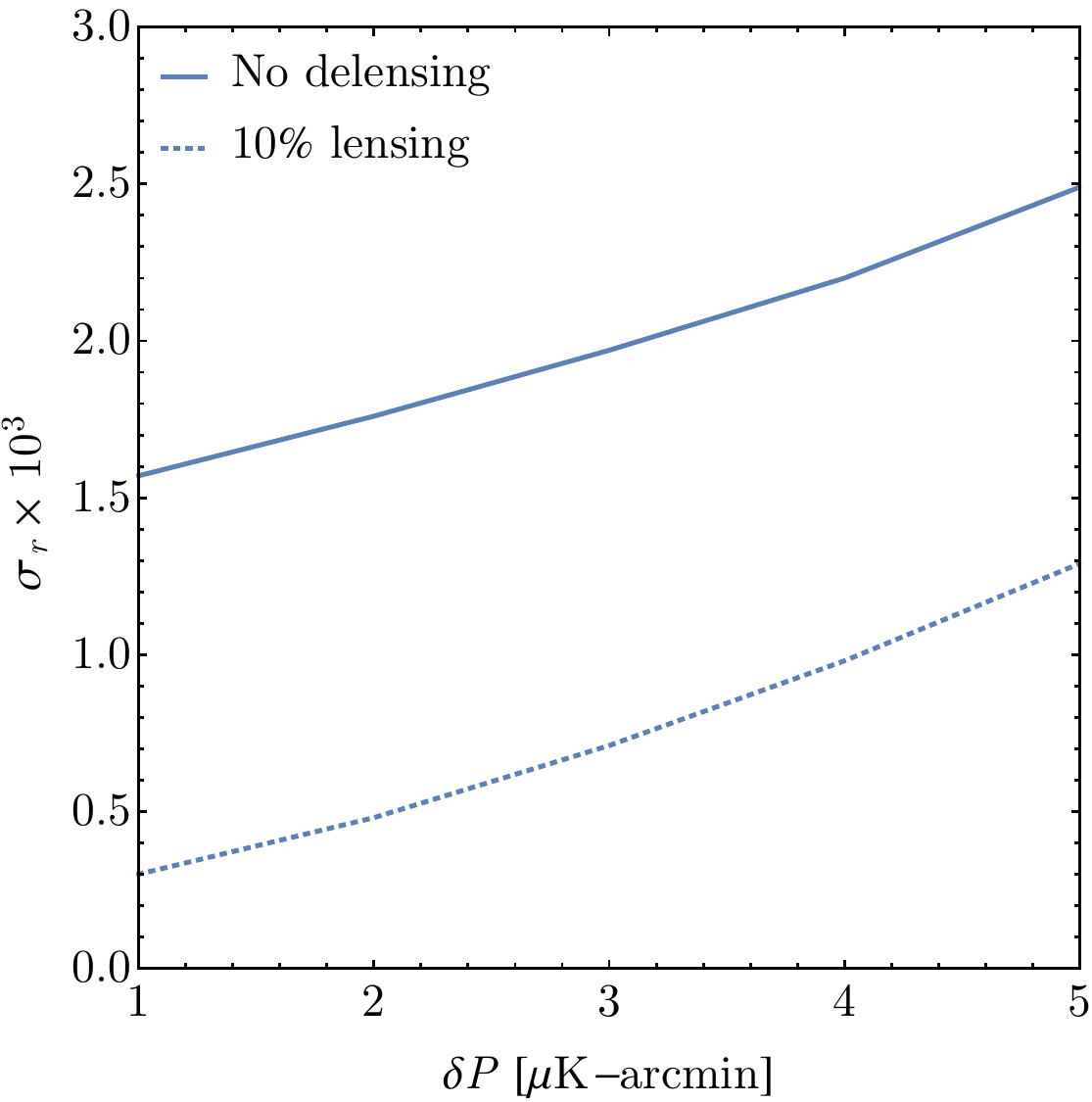}
        \caption{BAL covering 5\% of the sky.}
    \end{subfigure}%
    ~ 
    \begin{subfigure}[t]{0.5\textwidth}
        \centering
        \includegraphics[width= \textwidth]{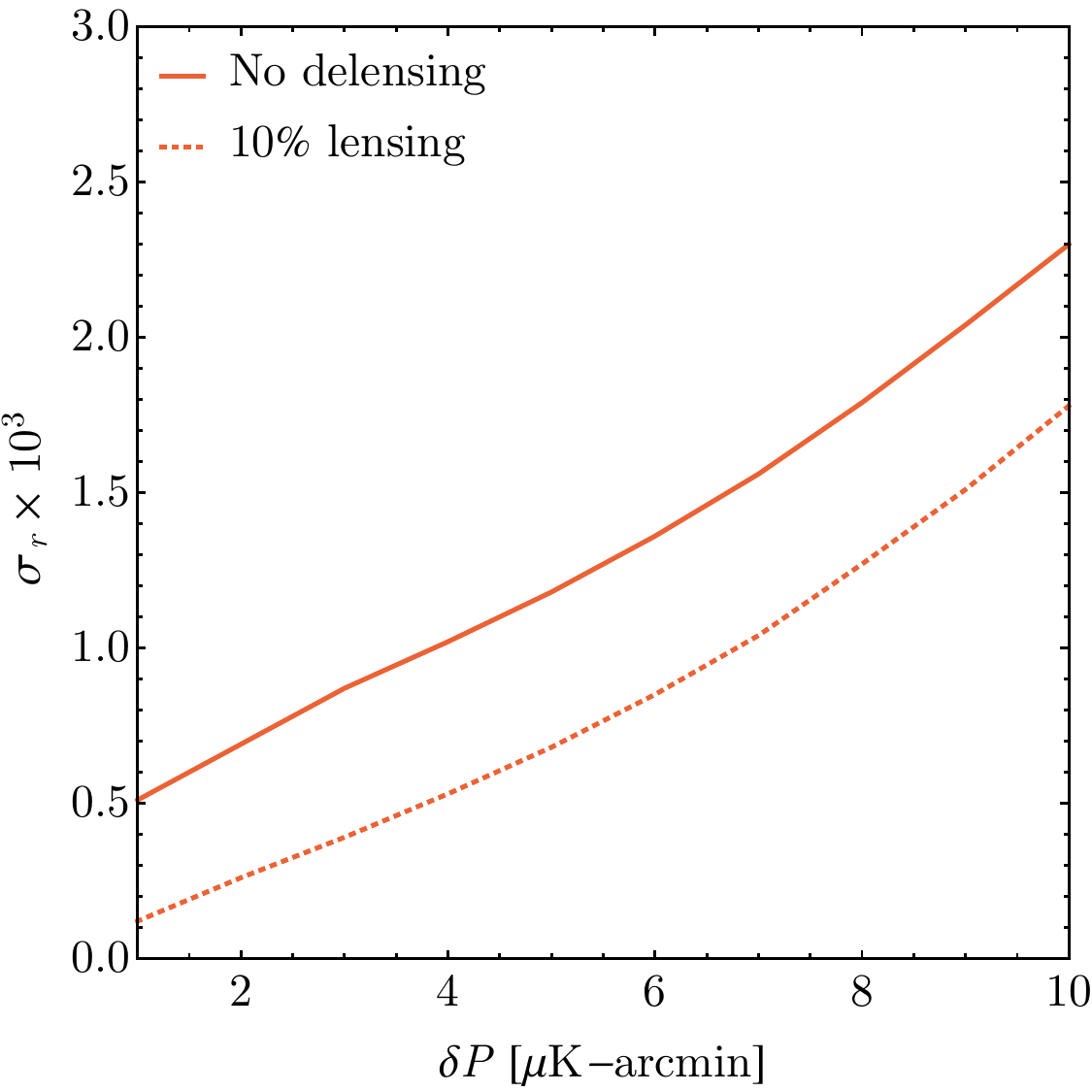}
        \caption{ULDB covering 60\% of the sky.}
    \end{subfigure}
    \caption{1$\sigma$ error on $r$ as a function of the instrumental sensitivity for two hypothetical balloon experiments (BAL and ULDB in Tab.~\ref{tab:SpecBAL}). These estimates use a patch of 5\% of the sky for BAL and 60\% of the sky for the ULDB. The solid line assumes lensing is not subtracted, the dotted line assumes lensing has been reduced to 10\% in power. Since also delensing is applied, we considered the multipoles [30, 300].}
\label{fig:BAL}
\end{figure*}


Regarding balloon-borne experiments, we consider EBEX 10k and Spider (which has already finished the first flight).  As can be seen in Tab.~\ref{tab:ErrorBAL}, we expect these experiments to explore values of $r\sim \rm{few}\times 10^{-2}$. It is fair to say that the level of dust measured by Planck only slightly degrade the previous forecasts and that the goal of these experiments are still within their range.

For all these experiments the error on the dust amplitude will be substantially smaller than the present Planck constraints, so that a cross-correlation with Planck will not significantly reduce the errors. However, Planck's data will still be useful to test the spectral dependence of the polarized dust emission model.

Looking a bit further into the future, we also consider an idealized balloon (BAL) and an ultra long duration ballon (ULDB) with the same four frequencies of EBEX 10k (150, 220, 280 and 350 GHz) and beams of 5 arcmin, but  leave their sensitivity as  a free parameter. For simplicity we assume that the sensitivity is equal across the frequencies, even though this may not be the optimal choice.  The results of our forecasts can be found in Fig.~\ref{fig:BAL} where we estimate the $1\sigma$ error for $r= 0$ for BAL covering the few supposedly clean patches found by Planck ($\sim5\%$ of the sky), and ULDB covering 60\% of the sky. As it can be seen, with a noise level $\sim 1\,\mu$K-arcmin and lensing removed to 10\%, which are possible but challenging, one can detect $r\sim 2\times 10^{-3}$ with high statistical significance. Notice however that obtaining a noise level close to $ 1\,\mu$K-arcmin on 60\% of the sky with an ULDB seems out of reach with a single 100-days flight (which is the target of this kind of experiments). 
We will discuss about a futuristic ground-based experiment (GRD) in Sec.~\ref{ResultsConservative}.\\


\subsection{Satellite Experiments} \label{ResultsSAT}
Finally, let us present our forecasts for various proposed satellite experiments (see Tab.~\ref{tab:SpecSAT} for the specifications). We assume that an effective area of 70\% is observed with foregrounds parameters given in Tab.~\ref{tab:NewForegrounds}, and limit the multipole range to $[2,300]$ for COrE and CMBPol (EPIC-2m) and $[2,150]$ for LiteBIRD\footnote{The angular resolution of LiteBIRD is not good enough for delensing and there is no advantage in considering higher multipoles when lensing is not subtracted.}. The summary of our estimates can be found in Tab.~\ref{tab:ErrorSAT}, and it shows that with respect to previous forecasts \cite{Baumann:2008aq} there is only a minor degradation (no more than a factor of 2) of the ability to detect primordial tensor modes. In particular a detection of $r\sim 2\times 10^{-3}$ is still achievable for the proposed missions. Notice however that the upper limit for $r=0$ below $10^{-4}$ are very optimistic and degrade significantly when the reionization bump is excluded (we are going to comment on this in the next section).
\begin{table}[htbp] 
\begin{center}
\begin{tabular}{ c c c c c }
\toprule
& {\bf r}  &{\bf CMBPol} & {\bf COrE}& {\bf LiteBIRD} \\
\midrule
\multicolumn{1}{ c }{\multirow{4}{*}{ CS} } 			& 0.1 		& $1.6\times 10^{-3}$	& $1.7\times 10^{-3}$	& $2.3\times 10^{-3}$\\
\multicolumn{1}{ c }{} 						& 0.01 		& $3.0\times 10^{-4}$	& $3.5\times 10^{-4}$	& $6.7\times 10^{-4}$\\
\multicolumn{1}{ c }{} 						& 0.001 	 	& $1.1\times 10^{-4}$	& $1.7\times 10^{-4}$	& $3.1\times 10^{-4}$\\
\multicolumn{1}{ c }{} 						& 0 			& $2.1\times 10^{-5}$	& $3.3\times 10^{-5}$	& $8.9\times 10^{-5}$\\
\midrule
\multicolumn{1}{ c }{\multirow{4}{*}{  FG 1\%}} 		& 0.1 		& $2.6 \times 10^{-3}$	& $2.6 \times 10^{-3}$	& $3.8 \times 10^{-3}$\\
\multicolumn{1}{ c }{} 						& 0.01 		& $5.7 \times 10^{-4}$ 	& $7.1 \times 10^{-4}$	& $1.3 \times 10^{-3}$\\
\multicolumn{1}{ c }{} 						& 0.001 		& $3.6 \times 10^{-4}$	& $5.1 \times 10^{-4}$	& $1.0 \times 10^{-3}$\\
\multicolumn{1}{ c }{} 						& 0 			& $3.4 \times 10^{-4}$	& $4.9 \times 10^{-4}$	& $9.9 \times 10^{-4}$\\
\bottomrule
\end{tabular}
\caption{ $1\sigma$ errors on $r$ for various proposed satellite experiments. For CMBPol and COrE  a delensing of 10\% has been taken into account.}
\label{tab:ErrorSAT}
\end{center}
\end{table}

\section{More conservative analyses}\label{ResultsConservative}

It is obvious, especially after the case of BICEP2, that a detection of primordial tensor modes must convincingly show that the signal is not contaminated by astrophysical foregrounds.
If the description of foregrounds in terms of few parameters is accurate, we saw that future experiments will be able to remove them with very good accuracy. On the other hand, our knowledge of astrophysical foregrounds is rather qualitative and it is not clear at what level the model works. For example, for $r = 2 \times 10^{-3}$ foregrounds at 150 GHz are larger than the primordial signal by a factor of 10 in amplitude at the recombination bump on the cleanest $1 \%$ of the sky, and a factor of 50 in the $70\%$ of the sky. 
\begin{table}[htbp] 
\begin{center}
\begin{tabular}{ c c c c c}
\toprule
& {\bf r} & {$\boldsymbol{ \sigma_r}$} & $\boldsymbol{ \sigma_{\alpha_{CMB}}}$ & $\boldsymbol{ \sigma_{\beta_{CMB}}}$ \\
\midrule
\multicolumn{1}{ c }{\multirow{2}{*}{{\bf AdvACT}}} 		& 0.1 	& $1.4\times 10^{-2}$	& $1.2\times 10^{-1}$	& $1.8\times 10^{-1}$	\\
\multicolumn{1}{ c }{} 							& 0.001 	& --- 					& ---					& ---\\
\midrule
\multicolumn{1}{ c }{\multirow{2}{*}{{\bf CLASS}}} 		& 0.1 	& $2.3\times 10^{-2}$	& $1.9\times 10^{-1}$	& $2.2\times 10^{-1}$ 		\\
\multicolumn{1}{ c }{} 							& 0.01 	& ---					& --- 					& ---	\\
\midrule
\multicolumn{1}{ c }{\multirow{2}{*}{{\bf Keck/BICEP3}}} 	& 0.1 	& $2.8\times 10^{-2}$ 	& $1.7\times 10^{-1}$	& $1.1\times 10^{-1}$\\
\multicolumn{1}{ c }{} 							& 0.01 	& --- 					& ---					& ---\\
\midrule
\multicolumn{1}{ c }{\multirow{2}{*}{{\bf Simons Array}}} 	& 0.1 	& $1.9\times 10^{-2}$	& $1.3\times 10^{-1}$	& $2.2\times 10^{-1}$\\
\multicolumn{1}{ c }{} 							& 0.01 	& ---					& ---					& ---\\
\midrule
\multicolumn{1}{ c }{\multirow{3}{*}{{\bf SPT-3G}}} 		& 0.1 	& $1.2\times 10^{-2}$	& $1.3\times 10^{-2}$	& $2.2\times 10^{-1}$\\
\multicolumn{1}{ c }{} 							& 0.01 	& $7.2\times 10^{-3}$	& $9.8\times 10^{-1}$	& 1.1\\
\multicolumn{1}{ c }{} 							& 0.001 	& ---					& --- 					& ---					\\
\midrule
\multicolumn{1}{ c }{\multirow{2}{*}{{\bf EBEX 10k}}} 		& 0.1 	& $1.6\times 10^{-2}$	& $4.7\times 10^{-1}$ 	& $3.9\times 10^{-1}$	\\
\multicolumn{1}{ c }{} 							& 0.01 	& ---					& --- 					& ---					\\
\midrule
\multicolumn{1}{ c }{\multirow{2}{*}{{\bf Spider}}}	 	& 0.1 	& $3.3\times 10^{-2}$ 	& $4.7\times 10^{-1}$	& $5.4\times 10^{-1}$\\
\multicolumn{1}{ c }{} 							& 0.01 	& --- 					& ---					& ---\\
\bottomrule
\end{tabular}
\caption{$1\,\sigma$ errors on $r$, $\alpha_{CMB}$ and $\beta_{CMB}$ for future ground-based and balloon-borne experiments.}
\label{tab:ErrorGRD10}
\end{center}
\end{table}

There are of course various ways to check that we are observing primordial gravitational waves. The primordial signal is homogeneous over the sky and it has Gaussian statistics, contrary to what we expect for foregrounds \cite{Kamionkowski:2014wza}. Other features that are well known about the signal are its dependence both in frequency and in $\ell$.
To study the ability of future experiments to check these features, we add to the parameters discussed in the previous section also the possibility of a power-law frequency dependence of the CMB signal $(\nu/\nu_{CMB})^{\beta_{CMB}}$ with $\nu_{CMB} = 150$ GHz. Moreover, we multiply the tensor mode power spectrum by a power-law $\ell$-dependence $(\ell/\ell_{CMB})^{\alpha_{CMB}}$ with $\ell_{CMB} = 80$. This roughly corresponds to the tensor tilt, although we are here interested in checking the expected approximate scale-invariance and not to assess the possibility to detect the tensor tilt. A convincing detection of primordial tensors should constrain both $\alpha_{CMB}$ and $\beta_{CMB}$ to be close to zero.  This will also give a sense of how close an unmodelled foreground component must be to the CMB signal to be undistinguishable from it. Since we want to be more conservative we also add as new parameters the $\ell$-dependence of dust and synchrotron ($\alpha_D$ and $\alpha_S$) so that the likelihood is a function of 10 parameters.
\begin{table}[htbp]
\begin{center}
\begin{tabular}{ c c c c c}
\toprule
& {\bf r} & {$\boldsymbol{ \sigma_r}$} & $\boldsymbol{ \sigma_{\alpha_{CMB}}}$ & $\boldsymbol{ \sigma_{\beta_{CMB}}}$ \\
\midrule
\multicolumn{1}{ c }{\multirow{3}{*}{{\bf CMBPol}}}	 	& 0.1 	& $1.6\times 10^{-3}$ 	& $2.2\times 10^{-2}$	& $6.8\times 10^{-3}$\\
\multicolumn{1}{ c }{} 							& 0.01 	& $3.9\times 10^{-4}$ 	& $5.1\times 10^{-2}$	& $2.6\times 10^{-2}$\\
\multicolumn{1}{ c }{} 							& 0.001 	& $2.1\times 10^{-4}$ 	& $1.4\times 10^{-1}$	& $1.3\times 10^{-1}$\\
\midrule
\multicolumn{1}{ c }{\multirow{3}{*}{{\bf COrE}}} 			& 0.1 	& $1.7\times 10^{-3}$	& $2.5\times 10^{-2}$ 	& $1.0\times 10^{-2}$	\\
\multicolumn{1}{ c }{} 							& 0.01 	& $4.9\times 10^{-4}$	& $6.1\times 10^{-2}$ 	& $4.3\times 10^{-2}$	\\
\multicolumn{1}{ c }{} 							& 0.001 	& $2.7\times 10^{-4}$	& $1.7\times 10^{-1}$ 	& $2.1\times 10^{-1}$	\\
\midrule
\multicolumn{1}{ c }{\multirow{3}{*}{{\bf LiteBIRD}}}	 	& 0.1 	& $2.7\times 10^{-3}$ 	& $3.9\times 10^{-2}$	& $1.5\times 10^{-2}$\\
\multicolumn{1}{ c }{} 							& 0.01 	& $1.2\times 10^{-3}$ 	& $1.2\times 10^{-1}$	& $7.3\times 10^{-2}$\\
\multicolumn{1}{ c }{} 							& 0.001 	& $8.0\times 10^{-4}$ 	& $3.4\times 10^{-1}$	& $3.7\times 10^{-1}$\\
\bottomrule
\end{tabular}
\caption{ $1\,\sigma$ errors on $r$, $\alpha_{CMB}$ and $\beta_{CMB}$ for future satellite experiments.}
\label{tab:ErrorSAT10}
\end{center}
\end{table}

The results for ground-based and balloon-borne experiments are reported in Tab.~\ref{tab:ErrorGRD10} and include only values of $r$ for which a significant detection is possible, since only in this case the additional parameters $\alpha_{CMB}$ and $\beta_{CMB}$ are relevant.
We see that the next generation of experiments will not be able to constrain $\alpha_{CMB}$ and $\beta_{CMB}$, unless $r \sim 0.1$.  

\begin{table}[htbp] 
\begin{center}
\begin{tabular}{ c c c c c c c c}
\toprule
& {\bf r} & {\bf Simons Array}& {\bf AdvACT}& {\bf CLASS}  & {\bf CMBPol}& {\bf COrE} & {\bf LiteBIRD} \\
\midrule
\multicolumn{1}{ c }{\multirow{4}{*}{CS}} 			& 0.1 	& $1.0\times 10^{-2}$	& $9.7\times 10^{-3}$	& $8.3\times 10^{-3}$	& $2.7\times 10^{-3}$	& $2.8\times 10^{-3}$	& $3.7\times 10^{-3}$\\
\multicolumn{1}{ c }{} 						& 0.01 	& $8.3\times 10^{-3}$	& $7.6\times 10^{-3}$ 	& $6.1\times 10^{-3}$	& $3.9\times 10^{-4}$	& $4.6\times 10^{-4}$	& $1.0\times 10^{-3}$\\
\multicolumn{1}{ c }{} 						& 0.001 	& ---					& ---				 	& ---					& $1.6\times 10^{-4}$	& $2.4\times 10^{-4}$	& $7.5\times 10^{-4}$\\
\multicolumn{1}{ c }{} 						& 0 		& $8.1\times 10^{-3}$	& $7.4\times 10^{-3}$ 	& $5.9\times 10^{-3}$	& $1.4\times 10^{-4}$	& $2.1\times 10^{-4}$	& $7.2\times 10^{-4}$\\
\bottomrule
\end{tabular}
\caption{ $1\sigma$ errors on $r$ for big-patch experiments, assuming $\ell>30$.}
\label{tab:ErrorNObump}
\end{center}
\end{table}

Our results for satellite experiments are given in Tab.~\ref{tab:ErrorSAT10}: we find that the inclusion of additional parameters does not significantly degrade the errors on $r$ (at most by a factor of 2) . Even for $r = 0.1$ the check of the tensor consistency relation, which would give $\alpha_{CMB} \simeq 10^{-2}$, looks impossible.

Another point of concern about foregrounds is the possibility of detecting the reionization bump. This of course is only relevant for experiments looking at a large portion of the sky. At this stage our knowledge of polarized foregrounds on large scales is very limited and it is not clear whether the reionization bump will be accessible once foregrounds are included. Moreover, ground-based experiments will also be limited by atmospheric contaminants. While in the previous sections we extended the analysis to low multipoles for experiments with large $f_{sky}$, in Tab.~\ref{tab:ErrorNObump} we consider a more conservative analysis where only the multipoles $\ell > 30$ are considered. To do so we consider the likelihood a function of six parameters, as in the previous section. While the change is moderate for large values of $r$, since the low multipoles do not help much with the statistics, the effect is relevant for small values of $r$ and becomes dramatic for $r=0$: the upper limit is degraded by a factor of 10 (this is compatible with the results of \cite{Smith:2010gu}).
Notice that the amplitude of the reionization bump depends strongly on $\tau$, so that when the measurement of tensor modes relies on the large scales, the error on $r$ is significantly affected by the uncertainty on $\tau$.

\begin{figure}[htbp]
\begin{center}
\includegraphics[width=0.55 \textwidth]{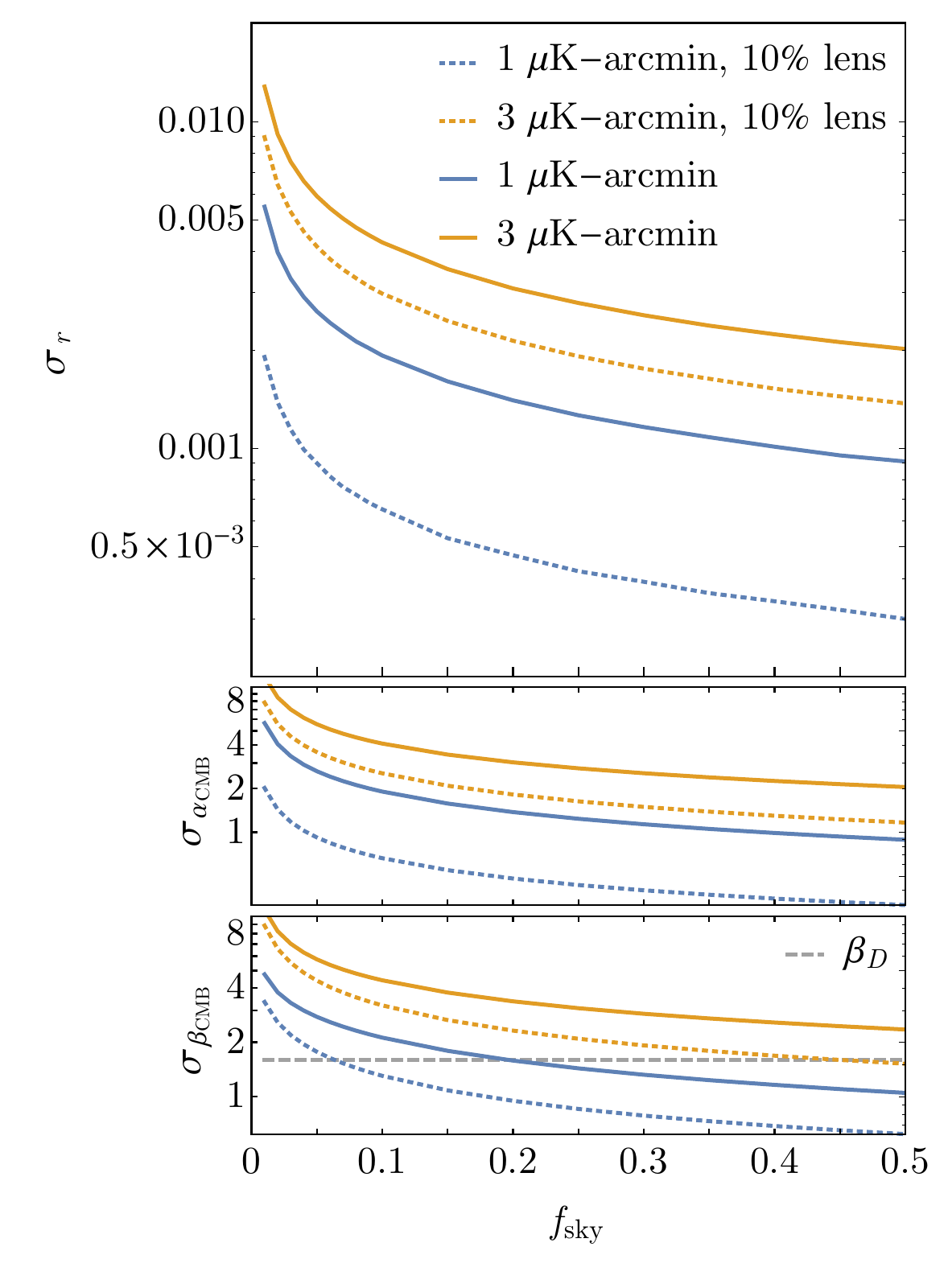}
\end{center}
\caption{\textbf{Top panel:} Error on $r=2\times10^{-3}$ as a function of $f_{sky}$ for a hypothetical ground experiment (GRD in Tab.~\ref{tab:SpecBAL}). Solid lines assumes lensing is not subtracted, dotted lines assumes lensing has been reduced to 10\%. Since also delensing is applied, we considered the multipoles [30, 300].\\ \textbf{Middle panel:} Error on $\alpha_{CMB}$.
\\ \textbf{Bottom panel:} Error on $\beta_{CMB}$. Notice that only for $f_{sky}\gtrsim30\%$ the constraints on the spectral dependence allows to reject dust at $2\sigma$.}
\label{fig:GRD}
\end{figure}

Let us now comment on our fiducial threshold $r = 2 \times 10^{-3}$. Ground-based experiments of the so-called stage IV are expected to achieve a sensitivity of the order $1\,\mu$K-arcmin with $\mathcal{O}(10^5)$ detectors over 5 years. For this value of $r$, in Fig.~\ref{fig:GRD} we show the error on $r$, $\alpha_{CMB}$ and $\beta_{CMB}$ for a hypothetical ground-based experiment (GRD) as a function of $f_{sky}$ for two different sensitivities. The detection of $r = 2\times 10^{-3}$ can be achieved at more than $3\sigma$ if the maps are delensed to 10\% and roughly 20\% of the sky is observed. In this case the constraints on $\alpha_{CMB}$ and $\beta_{CMB}$ are small enough to allow a clear distinction from our modeled foregrounds. For satellite experiments, from our results shown in Tab.~\ref{tab:ErrorSAT10} and \ref{tab:ErrorNObump}, we see that $r = 2 \times 10^{-3}$ is still detectable with large significance, even when it is not possible to detect the reionization bump. From Tab.~\ref{tab:ErrorSAT10} we also see that the error on $\beta_{CMB}$ is small enough to allow for a clear distinction of $\beta_{CMB}$ from $\beta_{D}$ (or  $\beta_{S}$).

The skeptical reader may be worried about the possibility of detecting gravitational waves buried under a foreground signal: we can model foregrounds, but how can we be sure that what is left in map is due to tensor modes and not some additional ``evil dust'' component we are unaware of? It is fair to say that a robust detection of primordial tensor modes requires a detection of the recombination bump. This feature, like a resonance in particle physics, should be robust against foregrounds which are not expected to peak al $\ell \sim 80$. To assess the ability of future experiments to measure the bump and distinguish it from a featureless power-law dependence, we compare the analysis in Sec.~\ref{Results} (extended to include $\alpha_{CMB}$), with a model which does not include the tensor transfer function, so that the spectrum is just a power law in $\ell$. By treating the amplitude of the bump as a continuos parameter, one can use Wilks's theorem.\footnote{We compute the minimum value of $r$ for which $\frac{1}{2}\left(1-\rm{CDF}\left(2\langle \log\mathcal{L}_{bump}\rangle-2\langle \log\mathcal{L}_{no\,bump}\rangle\right)\right)<0.003$, which corresponds to a 3$\sigma$ confidence level, where CDF is the cumulative distribution function for the $\chi^2$ distribution with one degree of freedom.}  In Tab.~\ref{tab:recombbump} we report for some of the experiments the minimum value of $r$ for which a $3 \sigma$ evidence of the bump (compared to a featureless power law $\ell$-dependence) is possible. The rule of thumb is that a $10 \sigma$ measurement of $r$ gives a $3 \sigma$ evidence of the recombination bump. For $r = 2 \times 10^{-3}$ it will be challenging to obtain evidence of the recombination bump even with satellite experiments. Indeed, for such a small value of $r$ the primordial power spectrum at the recombination peak is comparable to the lensing $B$-modes reduced to 10\%  in power or 1.4 $\mu$K-arcmin. 

\begin{table}[htbp] 
\begin{center}
\begin{tabular}{ c c c c c c}
\toprule
&{\bf Simons Array}& {\bf AdvACT} & {\bf CLASS} & {\bf GRD}  & {\bf CMBPol} \\
\midrule
$r_{min}$& 0.08 	& 0.055 & 0.095 & $0.005$  & $0.003$ \\
\bottomrule
\end{tabular}
\caption{Minimum value of $r$ for which a $3\sigma$ detection of the recombination bump is possible. For GRD we choose a noise of $1\,\mu${\rm K-arcmin}, $10\%$ delensing and $20\%$ of the sky. Since we are interested in the recombination bump, the analysis is restricted to $\ell > 30$.}
\label{tab:recombbump}
\end{center}
\end{table}

\section{Conclusions}
In this paper we updated the forecasts for various future $B$-mode experiments taking into account Planck data on foregrounds. For experiments with at least three frequencies, the forecasts on $r$ do not change significantly with respect to previous estimates, provided a simple modeling of foregrounds in terms of few parameters works at the required accuracy. 

In particular we focussed on the theoretically motivated target of $r = 2 \times 10^{-3}$. This is achievable both with balloon-borne and ground based experiments if the noise can be reduced to $\sim1\,\mu$K-arcmin and lensing $B$-modes are reduced to $10\%$. The ground-based experiments covering $\gtrsim 30\%$ of the sky should also have the statistical significance to check that the gravitational wave signal has a frequency dependence compatible with the one of the CMB and very different from the known foregrounds.  Even for satellite experiments observing the recombination bump, which would likely be a convincing evidence that the signal is indeed due to primordial tensor modes, will be challenging  for $r = 2 \times 10^{-3}$.

\subsection*{Acknowledgements}
It is a pleasure to thank C. Baccigalupi, D. Gaggero and F. Paci for useful discussions. We would also like to thank J. Errard and G. Fabbian for pointing out the new specifications of the Simons Array. We also would like to thank the unknown referee for the careful review of our paper. M.S. gratefully acknowledges support from the Institute for Advanced Study. M.Z. is supported in part by NSF Grants No. PHY-1213563 and No. AST-1409709.
\clearpage
\appendix
\section{Instrumental Specifications}
\begin{table}[H]
\begin{center}
\begin{tabular}{lcccc}
\toprule%
{\bf Experiment}  & $\boldsymbol{f_{sky}\, [\%]}$ &{\bf $\boldsymbol{\nu}$\, [GHz]} & $\boldsymbol{\theta_{FWHM}\, [']}$ & $\boldsymbol{\delta P\, [\mu \rm{K}']}$\\

\midrule
\multicolumn{1}{l}{\multirow{3}{*}{AdvACT}} 	&\multicolumn{1}{c}{\multirow{3}{*}{50}}	& 90 		& 2.2 	& 11 \\
\multicolumn{1}{l}{} 						& \multicolumn{1}{c}{}				& 150 	& 1.3		& 9.8\\
\multicolumn{1}{l}{} 						& \multicolumn{1}{c}{}				& 230	& 0.9		& 35\\
\midrule
\multicolumn{1}{l}{\multirow{4}{*}{CLASS}} 	&\multicolumn{1}{c}{\multirow{4}{*}{70}}	& 38 		& 90 		& 39 \\
\multicolumn{1}{l}{} 						& \multicolumn{1}{c}{}				& 93 		& 40		& 13\\
\multicolumn{1}{l}{} 						& \multicolumn{1}{c}{}				& 148 	& 24		& 15\\
\multicolumn{1}{l}{} 						& \multicolumn{1}{c}{}				& 217	& 18		& 43\\
\midrule
\multicolumn{1}{l}{\multirow{4}{*}{EBEX 10k}} 	&\multicolumn{1}{c}{\multirow{4}{*}{2}}	& 150 	& 6.6 	&  5.5\\
\multicolumn{1}{l}{} 						& \multicolumn{1}{c}{}				& 220 	& 4.7		& 11\\
\multicolumn{1}{l}{} 						& \multicolumn{1}{c}{}				& 280 	& 3.9		& 25\\
\multicolumn{1}{l}{} 						& \multicolumn{1}{c}{}				& 350	& 3.3		& 52\\
\midrule
\multicolumn{1}{l}{\multirow{3}{*}{Keck/BICEP3}} 	&\multicolumn{1}{c}{\multirow{3}{*}{1}}	& 95 		& 30 		& 3.0 \\
\multicolumn{1}{l}{} 						& \multicolumn{1}{c}{}				& 150 	& 30		& 3.0\\
\multicolumn{1}{l}{} 						& \multicolumn{1}{c}{}				& 220	& 30		& 10\\
\midrule
\multicolumn{1}{l}{\multirow{3}{*}{Simons Array}} &\multicolumn{1}{c}{\multirow{3}{*}{65}}	& 95 		& 5.2 	& 13.9 \\
\multicolumn{1}{l}{} 						& \multicolumn{1}{c}{}				& 150 	& 3.5		& 11.4\\
\multicolumn{1}{l}{} 						& \multicolumn{1}{c}{}				& 220	& 2.7		& 30.1\\
\midrule
\multicolumn{1}{l}{\multirow{3}{*}{Spider}}		&\multicolumn{1}{c}{\multirow{3}{*}{7.5}}  	& 94 		& 49 		& 17.8 \\
\multicolumn{1}{l}{} 						& \multicolumn{1}{c}{}				& 150 	& 30		& 13.6\\
\multicolumn{1}{l}{} 						& \multicolumn{1}{c}{}				& 280	& 17		& 52.6\\
\midrule
\multicolumn{1}{l}{\multirow{3}{*}{SPT-3G}} 	&\multicolumn{1}{c}{\multirow{3}{*}{6}}	& 95 		& 1 		& 6.0 \\
\multicolumn{1}{l}{} 						& \multicolumn{1}{c}{}				& 150 	& 1		& 3.5\\
\multicolumn{1}{l}{} 						& \multicolumn{1}{c}{}				& 220	& 1		& 6.0\\
\midrule
\multicolumn{1}{l}{\multirow{1}{*}{BAL}} 		& \multicolumn{1}{c}{5}				& 150, 220, 280, 350 	& 5 		& [1,5] \\
\midrule
\multicolumn{1}{l}{\multirow{1}{*}{ULDB}} 		& \multicolumn{1}{c}{60}				& 150, 220, 280, 350 	& 5 		& [1,10] \\
\midrule
\multicolumn{1}{l}{\multirow{1}{*}{GRD}} 		& \multicolumn{1}{c}{[1,50]}				& 100, 150, 220 	& 5 		& 1, 3 \\
\bottomrule
\end{tabular}
\caption{Specifications for balloon-borne and ground-based experiments used in our forecasts: \cite{Calabrese:2014gwa,ebex10k,2012SPIE.8452E..1AO,polar:spec,doi:10.1117/12.2057332,Fraisse:2011xz,Rahlin:2014rja,2014SPIE.9153E..1PB}. The sensitivity $\delta P=\sigma_{pix}\theta_{FWHM}$ is for the Stokes $Q$ and $U$. }
\label{tab:SpecBAL}
\end{center}
\end{table}
\clearpage
\begin{table}[H]
\begin{center}
\begin{tabular}{lccc}
\toprule%
{\bf Experiment} & $\boldsymbol{\nu\, [\rm{GHz}]}$ & $\boldsymbol{\theta_{FWHM}\, [']}$ & $\boldsymbol{\delta P\, [\mu \rm{K}']}$\\
\midrule
\multicolumn{1}{l}{\multirow{7}{*}{CMBPol (EPIC-2m)}} 	& 30 		& 26 		& 19.2 \\
\multicolumn{1}{l}{} 						& 45 		& 17		& 8.3\\
\multicolumn{1}{l}{} 						& 70		& 11		& 4.2\\
\multicolumn{1}{l}{} 						& 100 	& 8		& 3.2\\
\multicolumn{1}{l}{} 						& 150 	& 5		& 3.1\\
\multicolumn{1}{l}{} 						& 220 	& 3.5		& 4.8\\
\multicolumn{1}{l}{} 						& 340 	& 2.3		& 21.6\\
\midrule
\multicolumn{1}{l}{\multirow{11}{*}{COrE}} 		& 45 		& 23 		& 9.1 \\
\multicolumn{1}{l}{} 						& 75 		& 14		& 4.7\\
\multicolumn{1}{l}{} 						& 105	& 10		& 4.6\\
\multicolumn{1}{l}{} 						& 135 	& 7.8		& 4.6\\
\multicolumn{1}{l}{} 						& 165 	& 6.4		& 4.6\\
\multicolumn{1}{l}{} 						& 195 	& 5.4		& 4.5\\
\multicolumn{1}{l}{} 						& 225 	& 4.7		& 4.6\\
\multicolumn{1}{l}{} 						& 255 	& 4.1		& 10.5\\
\multicolumn{1}{l}{} 						& 285 	& 3.7		& 17.4\\
\multicolumn{1}{l}{} 						& 315 	& 3.3		& 46.6\\
\multicolumn{1}{l}{} 						& 375 	& 2.8		& 119\\
\midrule 
\multicolumn{1}{l}{\multirow{6}{*}{LiteBIRD}} 	& 60 		&32 		& 10.3 \\
\multicolumn{1}{l}{} 						& 78		& 58		& 6.5\\
\multicolumn{1}{l}{} 						& 100 	& 45		& 4.7\\
\multicolumn{1}{l}{} 						& 140 	& 32		& 3.7\\
\multicolumn{1}{l}{} 						& 195 	& 24		& 3.1\\
\multicolumn{1}{l}{} 						& 280 	& 16		& 3.8\\
\bottomrule
\end{tabular}
\caption{Specifications for satellite experiments used in our forecasts: \cite{Baumann:2008aq,Bouchet:2011ck,2014JLTP..176..733M}. The sensitivity $\delta P= \sigma_{pix}\theta_{FWHM}$ is for the Stokes $Q$ and $U$. All experiments target approximately 70\% of the sky.}
\label{tab:SpecSAT}
\end{center}
\end{table}

\printbibliography
\end{document}